# Curbing cyber-crime and Enhancing e-commerce security with Digital Forensics

Israel D. Fianyi
[1] School of Computing and Mathematics, Charles Sturt University, Australia

**Abstract**

The explosion in the e-commerce industry which has been necessitated by the growth and advance expansion of Information technology and its related facilities in recent years have been met with adverse security issues thus affecting the industry and the entire online activities. This paper exams the prevailing security threats e-commerce is facing which is predominantly known as cyber-crime and how computer related technology and facilities such as digital forensics can be adopted extensively to ensure security in online related business activities. This paper investigated the risk, damage and the cost cyber-crime poses to individuals and organizations when they become victims, since transacting business online as well as all related online activities has become inherent in our everyday life. It also comprehensively elucidate on some of the cyber-crime activities that are posing serious threat to the security of E-commerce. Amazon and EBay were used as the case of study in relation to respondents who patronizes these renowned e-commerce sites for various transactions.

## 1. Introduction

The emergence and popularization of the world wide web (WWW) has facilitated a drastic explosion in E-commerce and consequently has become an essential and indispensable constituent of business strategy and operations, hence the recent advancement in businesses, thus the main medium for trade and industry development in any given country and the world at large (Guan, Tan, & Hua, 2004). Due to this rapid growth in the e-commerce industry, many computer programmers and Information Security experts have directed their abilities to the development of applications that enhances computer assisted transactions via the internet. Smith (2009) indicated that the current revolution that has overtaken the world or business has become successful as a result of the incorporation of ICT in almost every facet of business transactions and its operations. It has also facilitated a strong correlation between firms and their corresponding clients. Particularly, the use of ICT in business has enhanced productivity, promoted customer involvement, also made faster and possible patronage of goods and services in larger quantities, as well as reducing charges and its associated benefits (Baek & Hong, 2003). In contrast to the normal brick and mortar business where there is physical contact between the business owner and the buyer. Online business is solely done via the internet without any form of contact as the transaction is done via a technology mediated platform.

Payment is done directly into the business owners account or through a third party payment system in this regard the customer is made to declare his/her confidential bank details to enable the other party get the money and then send whatever has been bought through a delivery service. There is virtually no physical encounter between the parties (Gnanasekar, 2014). Despite the successes e-commerce is experiencing in the e- world, there remain an issue of security which most often is to the detriment of the online shopper and consequently the business owners as well. The upsurges in cyber-crime by unknown assailant who usually take the advantage the internet offers and in anonymity perpetrate heinous crime under the notion that they cannot be seen, most crimes are perpetrated with this belief system. (Perez, 2005). There are various crimes that are committed via the internet and its related technologies from denial of services, identity theft, Privacy violation and intrusion as well as industrial and financial espionage, this study expounds on these crimes further in the following sections and in appendix 1. Perpetrators of these crimes do it for various reasons with the underlining fact of illegally making money for themselves among other reason (Sidel, 2005).

There are various factors that has propelled the security threat being experienced in recent times. However one irrefutable reason is the advancement in hacking tools (Andam, 2003). Hackers now capitalizes on the growth of technology to develop sophisticated applications to commit various crimes.

While most normal firms have established procedures for solving the common problems that comes with doing business, many online merchants and their customers do not possess the experience and technical know-how and usually not sure how to best find solutions to the critical and technical issues involved in cyber-crime detection (Abels, Brehm, Hahn, & Gomez, 2006). For this reason, there is the need for pragmatic steps to establish policies and technical know-how to combat this risk associated with online business and e-commerce at large. This goes a long way to save the business from embarrassment, loss of finances, legal challenges among other (Alsaad, Mohamad, & Ismail, 2014).

This paper investigated the measures that are undertaken to detect and prevent cyber-attack with digital forensic tools with regards to assisting customers to reduce exposure to fraudulent transactions, thereby minimizing fraud-related business losses (Aiken, Garner, Ghosh, & Vanjani, 2014) . By employing advanced modern ICT tool such as digital forensics by a firm which have the necessary mechanisms to investigate, curb cyber-crime and uncover perpetrators stand the chance of curbing e-crime. In the bid of most businesses ensuring security in their own prowess have set-up these systems that underpin the immediate first hand security in a respective organization: (a.) Setting up Clear Policies (b) Resolving disagreements (c) Systems that protect customers debit or credit card and /or any other medium used in making payment online (d) How to use digital forensics and related tools to detect and prevent cyber fraud (e) Examine the pay flow fraud prevention services

## 2. E-commerce and E-Risk

Risk is an eminent constituent of any business operation in any part of the world, e-commerce in this respect is not immune to risk associated with the internet and its related platforms, which most researchers call e-risk denoting the uncertainty of transacting business online. Especially in recent revolution of mobile commerce which has enabled businesses to rapidly expand (Kim, Hong, & You, 2015) .Some of the risk that are associated with e-commerce includes but not limited to, stolen Credit card number, personality theft, stolen Social Security Numbers, and sometimes goods bought and paid for never arrives. All these eminent issues and more are rampant because there is no physical encounter between the buyer and the seller basically; electronic mediums as mentioned earlier are employed as cited in Williams (2014). The research further reviews a statistics by cyber source a provider of electronic imbursement and threat control solutions, estimates that $3 billion was missing as a result of the above mentioned anomalies in e-commerce between 2006 and 2009 in the United States (Trautman, 2014).

Systems Security threats have progressively increased from both outside and within corporate boundaries. Internal security threat used to be the challenge of most corporate firms, these threats characteristically included employees within a company getting hold of company systems through loopholes in security and also having access to passwords they are not supposed to. There have been marginal changes from internal security threat to external threats due to the growth of the Internet and corporate extranets. Intruders from outside corporate boundaries now represent a growing concern (O'Brien & Marakas, 2011)

Since online activities have become imperative and indispensable to the daily lives of individuals and businesses, Security tools such as digital and computer forensic that are meant to investigate and ensure security on the web and its interrelated technology needs to be enhanced in its capability to withstand the growing security threat that confront E-commerce (Achille & Roger, 2014).The authors O'Brien and Marakas (2011) further established that imposters in cyberspace look for opportunity to penetrate and abuse systems, these they do either for the fun or selfish gains amidst other reasons. Once system is infiltrated, shams can possibly cause devastating problems such as deleting or altering information, illegally take valuable data as well as money in huge sums in that context .

A research by Turban, King, Lee, Liang, and Turban (2015) categorized some of the reasons that makes e-commerce susceptible to risk and they include the following;
(a) Old methods of solving e-business problems may not work due to changing environment.
(b)The more an e-commerce company expands the higher the scope of risks.
(c) System hackers will always devise new techniques and tools to attack system operations.
(d) When a system is digitized, it creates exceptional challenges for digital information and online transactions.

E-commerce risk and security is somehow known to the prospective customers, as mentioned by LaRose and Rifon (2007). The Internet though presents many opportunities to companies and individuals, the benefits cannot usually be compared to the degree of damage that results from security breaches. Hackers, phishing, identity theft, credit card details theft, viruses, malware, among others are major Security issues that confronts and frightens individual customers and companies at any level of online related business or E-commerce. These challenges used to be often resolved by the use of Hardware's and software's techniques to guard against hackers and infectious viruses from their intranet systems (Moore, 2010). Security goes beyond Hardware and software systems; they only represent one layer of security. Security as it stand will continually be a concern for both big and small firms and the lack of it can be devastating to the business due to the cost of resolving security problems, the possibility of huge financial loses and eventually lose of customers.

In contrast to traditional crimes where there are well established legal jurisdiction across all those crimes for prosecution, it is not the case with online crime that are perpetrated, there are some of the crimes that yet to be established as a crime in the legal environment. This differs from country to country and it makes it more challenging to subject some of these crimes committed via the World Wide Web to legal interpretation (CIOLAN, 2014; Vaseashta, Susmann, & Braman, 2014). For cyber related crime investigation to be successful with its corresponding collaring and prosecution an addendum should be made to the existing perceptions, legal barriers, and the procedures employed by law enforcement agencies to combat cyber-crime.

It is in this light that automated software's such as digital forensics which has seen revolution in recent years in crime investigative environment. Though it is been used its algorithms structures and capabilities have been undermined due to the fact that most law enforcement officers lack the expertise to digitally analyzed crimes committed in the this digitized world and as such present it in court as an evidence (Elyas, Ahmad, Maynard, & Lonie, 2015).

## 3. Digital Forensics

The evolution of Digital forensics which primarily focuses on extraction of data and analyzing crimes scenes and equipment after they have been committed and present them in court as an evidence, has in recent times become a necessary tools and software's applications in the fight against . Thus modern techniques of digital forensics , which can easily track a premeditated crime over the network by employing network security phenomenon to impede these possible crimes from occurring, can greatly be explored to help control the upsurge of e-commerce crime in recent times (Mylonas, Meletiadis, Mitrou, & Gritzalis, 2013). Subsequently with digital forensics tools crimes could be detected and prevented before they are eventually committed.

When forensics is mentioned it usually brings to mind television and movie dramas, emphasizes on murder scenes or a procedure involving a specialist surgeon. Nevertheless, a growing concern of technology demands retrieving information from a computer system have given another phase to forensic sciences which takes into consideration all investigative analysis that has to do with technologies. contemporarily digital forensic analyst may work along with law enforcement institutions; retrieving hidden information from a home computer system and office computer, mobiles phones and other digital gadget is very common now (Lochner & Zinn, 2014); These information may range from deleted files to existing documents stored on a hard drive, Network activities or other storage media and also encrypting files that may have been used in committing online related crimes (Stephenson & Gilbert, 2013).

3.1 Constitute Cybercrime

There are various crimes that are perpetrated via the internet by reason of the following category (Avzu, 2014)
• Computer assisted crimes; this is where the computer is used as a tool to enhance a criminal activity and these activities are not exclusive to computers only, example will be child pornography, fraud, cyber bully etc.
• Computer Specific crimes; these crimes includes password attacking, server attacks, system hacking, denial of services, sniffers. These crimes are directed to the computers and available digital gadget, the network and the various systems on which data is stored.
•Computer incidental; The computer becomes supplementary to crime perpetration such client list on spreadsheets, drug traffickers and terrorist interchanging information on various apps on the computer such as Skype, Facebook and the like.

The aftermath of a cyber-crime has varying effects usually in a passage of time. A typical example is about computer viruses which use to be seen as very dangerous and damaging to the any computer system that gets infected, in recent years it is not that dangerous as it used to be because stronger anti-viruses have come up to reduce the effects of damage they can cause to a system (Sommer & Paxson, 2010). In most of the cases carried out on cyber-crime, the person responsible is a hacker, or has relatively some computer expertise. Cyber-crime includes cyber-terrorism, electronic-theft, espionage, denial of service (DOS) credit card fraud, and phishing. Damages survived were closure of websites to stolen classified information (Broucek & Turner, 2004). Also there were legislations across the globe against computer virus proponent, it was seen as a criminal offence tantamount to prosecution, and this also assisted in reducing viral infections to systems. The sharing of information between local agencies, Federal states exposed possible virus attack a typical example is the collaboration that exist between the Federal Bureau of investigation and the secret service, Law enforcement agencies in in most developed countries (Hallahan, 2010).

These collaborations when encouraged by sharing developed digital forensics tools and other system security applications meant to curb cybercrime will go a long way to enhance a degree of security in a respective system, eventually cyber terrorist will be brought to book.

3.2 Supplementary Threats to Computer Security

Though unlawful access incapacitating a system has severe complications, they don't constitute the main threat to computer security. There are basically five threats that undermines a digitized system (Hartono, Holsapple, Kim, Na, & Simpson, 2014): (a) natural catastrophes, (b) corrupt workforces, (c) unhappy workforces, (d) and inadvertent blunders and oversights.

Human errors constitute the major cause of the breaches that occur in computer security, inadvertent oversights and mistakes are very common in the implementation and operation process of a system .However if a system is well developed these continues errors will be reduced hence the need for an effective internal control measures (Ghobakhloo, Arias-Aranda, & Benitez-Amado, 2011).

According to surveys carried out, showed that systems development is very effective when users are part of the process, the system will not likely fail. The following steps by organizations are essential to effective computer security. Security policies and controls should be enforced and communicated constantly (McLeod & MacDonell, 2011).
(a) Controls and security techniques should be put in place to ensure access to users and traceable information system should be implemented
(b) Users should be restricted to their area of jurisdiction only and not to other part of the system.
(c) Intermittent security training should be conducted.
(d) Some personnel should be made to be responsible for the security of the system in a liberated manner.

3.3. Network Forensic

Defensive measures can be engaged to ensure the prevention of cyber-crime. Nonetheless, irrespective of how many preventive measures are in place, except accurate and constant arranged procedures are maintained, other than that the system may either negligibly report cyber-crimes or excessively as a result giving false alarm, by a single interruption detection (Digital evidence, 2000). Interruption discovery methods include detection of anomaly in the system, tripwires, and configuration checking tools. This is possible by the use of forensic tools that solely investigate real-time online activity.

Network forensic is the branch of digital forensics that is characterized by capture, documenting and analyzing of activities that take place on a network for the purposes of identifying imposition and subjecting them to investigation. The concept of forensic is classified into two sections, one that enhances the security of the network and the other looks for anomalies on the network to establish evidence(Pilli, Joshi, & Niyogi, 2010).There are so many tools that underpins these network forensics from TCPDump, Flow-tools, IOS NetFlow, TCPXtract , and Argus among others. These areas of digital forensic are some of the directions that security experts can employ to curb e-commerce crime. Where every transaction that is taking place over the network is being monitored and red flag raised over the slightest alarm of possible crime perpetration. These Prevention techniques may not really be without some form of imperfection, as such, e-commerce firms should establish measures for analysis and revitalization the system after cybercrime attack occurs (Kaushik, Pilli, & Joshi, 2010).This will enable the firm to be guarded against subsequent infiltrations.

Most e-business firms usually require competent and experienced computer security persons; thus, hiring outside experts, for example, forensic accountants, computer forensic professional. If negligence of a company's computer security personnel caused a security breach or electronic crime, a professional from outside will still be needed and useful to examine the entire security of a company's system. Though many law enforcement bodies do not have the requisite expertise to investigate some issues of cybercrime, periodic training would enable them to understand the concept of digital forensics, this facilitates smooth corroboration of cyber-crime evidence discovered by the forensic specialist (Park, 2011).

Digital forensics s is concerned about an investigation into anything that has technology connotation, just like forensic science, it subject systems and technologies to observation and investigate the technology against anything that may be of the interest of the investigator. Technology has advanced and computer programmers have applications that can track criminal irrespective of their location by making use of the GPS. If digital forensic is introduced into e-commerce at an advanced level, it will reduce if not totally eradicate cyber-crime. This implies that business owners can determine the location of a shopper or customer if the person happens to be a hacker he/she can easily be identified (Vaseashta et al., 2014).

Authentication, authorization, and encryption are some basic security methodologies; the policy and privacy issues that governs e-commerce should be made available to prospective customers in various ways. It is imperative that e-commerce firms do not play on the ignorance on consumers in any given situation (Turban et al., 2015).

The law should be reconsidered to accept the scientific and procedural discoveries of digital forensics in its attempt to detect and expose perpetrators of cyber-crime. Observation of ethics in the practice to unravel criminal activities over the internet is as equally important as catching the criminals. According to Moore (2006) the court will usually accept digital forensic evidence in a typical cyber fraud case if the investigation and analysis was done within a generally accepted scientific parameters and ethical considerations not compromised.

## 4. Method

The respondents in this study were customers who buy from the internet particularly the most common ones Amazon and EBay thus they were the organization considered for the study. A sample size of 60 respondents (N=60) were randomly selected to participate in the research and there were 100% responses, no allurement to participants, it was purely voluntary responses.

## 5. Results

A descriptive analysis techniques was used to examine the data. The collected questionnaires were numbered to avoid duplicate data entry. The data was then computed into a SPSS, a software for data analysis. Each written answers of the respondent were coded under various headings as seen in the discussion below. A frequency distribution with its corresponding percent of the results to the questions were displayed

Table 1: Table 1. How respondents paid for items online

| Payment type | Frequency | Percent (%) |
|---|---|---|
| Visa Debit card | 21 | 35 |
| Visa Credit card | 15 | 25 |
| Master Card | 13 | 21.7 |
| American express | 11 | 18.3 |
| Total | 60 | 100 |

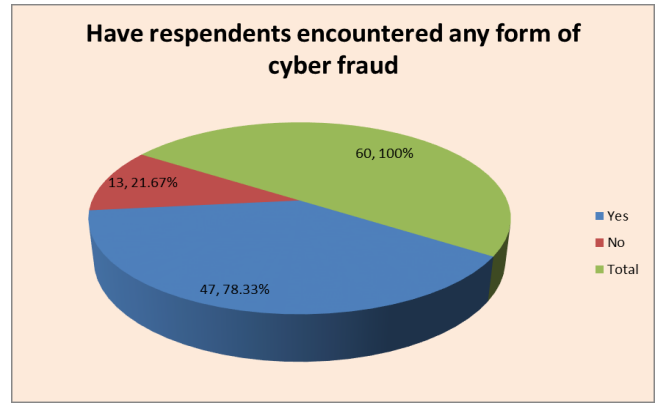

Fig. 1 Responses on respondents encounter of any form of fraud, delays in online transaction

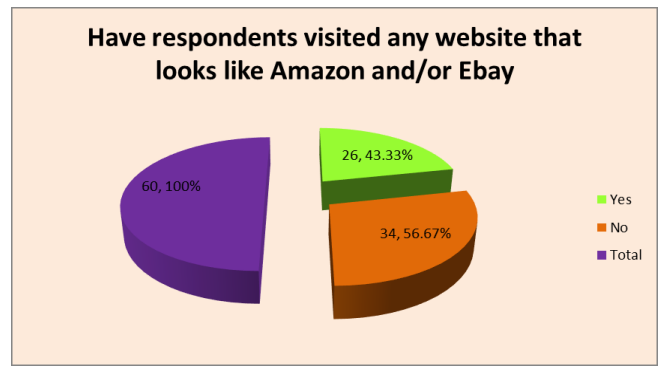

Fig. 2 Respondents visiting the wrong sites unknowingly

## 4. Conclusions

For the purpose of this paper three (3) of the findings are displayed. Firstly it saw that majority respondents paid for their items they purchased online with a visa debit card, thus these funds are sitting in their account, this confirms a research by Antonucci (2014), that easy way of transacting or buying online is to use a debit card as most firms prefer this option to others. The least reported medium of payment online was the use of the American express which suggest, not many people use it.

The use of debit and credit card for payment in recent times is a global phenomenon. Over decades now these cards have a magnetic hoop attached to it on these stripes are the relevant accounts details of the bearer. A more advanced type which was also recently introduced known as the integrated circuit thus IC which has chip and memory can be used to engage in a transaction with a contactless or contact systems, usually traders have point of sale stations and card readers that allows the use of card for payment of products. When a card is used at a sale terminal when the holder is physically present it is usually recorded in the bank transactions as card present as such there is less probability of fraud in situations like this because the holder is made to enter a password to authenticate ownership and processing of the transaction(Williams, 2014).

On the other hand, card not present is used on the description of transactions made typically over the internet and card details were used to finalize the payment. In a normal e-commerce transaction the cardholder communicate with the businesses' website where payment is made for chosen products and finally choses payment option either by use of Visa® card American express® or MasterCard® and so on. Besides the use of the card other details of the card holder are collected such as billing address and the three digit card verification value or code. This risk involved in this card not present transaction usually made over the internet is very high as far as the card holder or the customer is concerned (Williams, 2014). It would really be appropriate that some form of chip and pin transactions are available for internet related transactions in that regards to ensure the security of the card owner. Secondly the inquiry saw that a wide number of e-commerce users have encountered internet related fraud one way or the other whiles the least have not experienced any yet, this finding also confirms the findings of Dwyer and Easteal (2013) that cybercrime is really on the increase and advanced counter measure is imperative. A survey carried out by Karake-Shalhoub and Al Qasimi (2007) indicated that there are about 63% levels of un authorized penetration in Saudi Arabia as such most people have been victims of cyber-crime and it is really on the increase since the use of the internet has become part of their daily activities.

The cyber-crime known as phishing which is one of the most common cyber-crime where sham e-commerce web sites are created to look like the original by imposters so that as customers enter their card details and other important information they illegally take them or where a company's website is high jacked by hackers known as denial of service attacks with a similar purpose among others, a typical example was when on 6th August , 2009, Facebook, twitter and Google blogger were bashed leaving these social networking sites struggling to recover. Most users were not able to have access other did the cost of lost to both these firms and innocent customers were enormous (Broucek & Turner, 2001). There is also a current one known as vishing due to the explosion of smart phones they are used one way or the other to illegally obtain information from unsuspecting individuals.

## Appendix

Types of digital related crimes

| | |
|---|---|
| Computer virus | Computer program that causes serious damage to computer systems or files |
| Phishing | when the perpetrator set up deceitful websites that appears official and causing the victim to give out personal information to the criminals |
| Vishing | Personal details stolen via phone |
| Botnet | this occurs when a hacker transmits instructions to other computers for the purpose of controlling them |
| Spoofing | its use of email to swindle an individual into providing personal information that is later used for unlawful purposes |
| E- theft | The perpetrator hacks into a financial institution e.g. a bank and diverts funds to accounts accessible to the criminal. To prevent e-theft, most major banks severely limit what clients can do online |
| Cyber terrorism | Cyber terrorism occurs when terrorists cause virtual destruction in online computer systems |
| Espionage | occurs when perpetrators hack into online systems or individual PCs to obtain confidential information for the purpose of selling it to other parties (criminals) |
| Malware | This is constituted by a category of viruses such as spyware, Trojan horse, worms with its original intent to break the host system |
| Spam | Refers to unsolicited email; spam is illegal if it violates the Can-Spam Act of 2003, such as by not giving recipients an opt-out method. |

Source: (Andam, 2003; Carol Mendelsohn, 2015 May 13)

**Author:** A full time PhD candidate at Charles Sturt University Australia in the school of Computing and mathematics, Obtained Bachelor of Art in Information studies 2007 from University of Ghana, MBA in Management Information Systems 2010, University of Ghana and MBA E-commerce in 2014 from University of Gloucestershire UK. All working experience have been in the academia, specifically in IT related fields. Currently researching into Biometric security, digital forensics and Big data analytics.